\documentclass[12pt,cite,a4paper]{article}

\usepackage{amsmath}
\usepackage{amssymb}
\usepackage{a4wide}
\usepackage{amssymb}
\usepackage{graphicx}

\newcommand{\cA}{\mathcal{A}}
\newcommand{\cB}{\mathcal{B}}

\numberwithin{equation}{section}

\begin{document}
\rightline{USTC-ICTS-08-06}
 \vspace{2truecm}

%%%%%%%%%%%%%%%%%
\centerline{\Large \bf Mass Deformation of the Multiple M2 Branes
Theory}

\vspace{1.3truecm}

\centerline{
    {Yushu Song ${}^{a,b}$}\footnote{yssong@itp.ac.cn}
    }
\vspace{.8cm} \centerline{{\it ${}^a$Interdisciplinary Center of
Theoretical Studies}} \centerline{{\it USTC, Hefei, Anhui 230026,
China}} \vspace{.4cm} \centerline{{\it ${}^b$ Institute of
Theoretical Physics}} \centerline{{\it Academia Sinica, Beijing
100080, China}}

\vspace{2.5truecm}

%%%%%%%%%%%%%%%%%
\centerline{\bf ABSTRACT}
\vspace{.5truecm}

Based on recent developments, in this letter we study the one
parameter deformation of $2+1$ dimensional gauge theories with scale
invariance and $\mathcal{N} = 8$ supersymmetry, which is expected to
be the field theory living on a stack of M2 branes. The deformed
gauge theory is defined by a Lagrangian and is based on an infinite
set of novel $3$-algebras constructed by relaxing the assumption
that the invariant metric is positive definite. Under the Higgs
mechanism, we can obtain the D-branes world volume theory in the
presence of background fluxes. \noindent

%%%%%%%%%%%%%%%%%%%%%%%%%%%%%%%%%%%%%%%%%%%%%%%%%%%%%%%%%
\newpage

\section{Introduction}

M-branes are mysterious objects and virtually little is known about
their underlying dynamics. This is in sharp contrast to D-branes,
where a microscopic description in terms of open strings has driven
a huge amount of progress in string theory and gauge theory. The
three-dimensional superconformal field theory which is supposed to
describe multiple coincident M2 branes may lead to profound new
insight in our understanding of M-theory. Recently Bagger and
Lambert \cite{Bagger:2006sk,Bagger:2007jr, Bagger:2007vi} and
Gustavsson \cite{Gustavsson:2007vu, Gustavsson:2008dy}, proposed a
new set of $2+1$ dimensional field theory (henceforth called the BLG
theory) which is supposed to describe low energy world volume theory
of multiple coincident M2 branes. The BLG theory was constructed in
frame of so-called 3-algebra, a generalization of the Lie algebra
with triple bracket replacing the commutator and the 4-index
structure constant replacing the usual 3-index structure constant of
the Lie algebra. There are two requirements of 3-algebra in the BLG
theory: one is called the fundamental identity which is the
generalization of the Jacobi identity of the Lie algebra; the other
is that the metric of 3-algebra is positive definite. Recently the
BLG theory was generalized to a novel 3-algebra by relaxing the
assumption that the metric on 3-algebra is positive definite
\cite{Gomis:2008uv}\cite{Benvenuti:2008bt}\cite{Ho:2008ei}.
Henceforth we will call this theory generalized BLG theory. The
generalized BLG theory has many features which suggest that it is
related to M2 branes. For example, the gauge interaction term is the
BF-type which do not admit a tunable coupling constant and this
property extends to the full generalized BLG theory. The key point
is that this construction starts from the arbitrary Lie algebra, so
that we can construct theory of arbitrary number of M2 branes which
makes the well-developed large N tool \cite{Maldacena:1997re}
possible in this field.

In this letter we construct the one parameter deformation of the
generalized BLG theory and then we use the strategy of
\cite{Mukhi:2008ux} to show the relation between the deformed M2
branes theory and the induced D2 branes system. The deformed of BLG
theory was considered in \cite{Bagger:2007vi} \cite {Gomis:2008cv}
\cite{Hosomichi:2008qk} . The rest of this letter is organized as
follows. In section 2, we give a brief review of the BLG theory and
its generalized form. In section 3, we consider the deformation of
the generalized BLG theory. In section 4, following Mukhi and
Papageorgakis, we consider the reduction of M branes to D branes in
the presence background fluxes. In section 5, we will give some
discussions. For other recent developments of the BLG theory, see
\cite{Bandres:2008vf}-\cite{Krishnan:2008zm}.

\vspace{0.5cm} \textbf {Note added}: After this letter was finished,
the preprint \cite{Honma:2008un} focusing on Janus field theory
appeared on arXiv with substantial overlap with our results.

%\newpage
%%%%%%%%%%%%%%%%%%%%%%%%%%%%%%%%%%%%%%%%%%%%%%%%%%%%%%%%%%%%%%%
%%%%%%%%%%%%%%%%%%%%%%%%%%%%%%%%%%%%%%%%%%%%%%%%%%%%%%%%%%%%%%%
\section{Brief Review of Generalized BLG Theory}
The BLG theory is based on 3-algebra, which is the generalization of
Lie algebra. A 3-algebra is a $N$ dimensional vector space with
basis $T^A\ (A=1,2,....N)$ which is endowed with a trilinear
antisymmetric product
\begin{equation}\label{algebra relation}
[T^A,T^B,T^C]={f^{ABC}}_{D}T^D
\end{equation}
where ${f^{ABC}}_D$ is the structure constant. From (\ref{algebra
relation}) it is clear that ${f^{ABC}}_D={f^{[ABC]}}_D$. Then
further suppose there is a trace form providing a metric
\begin{equation}
h^{AB}={\rm Tr}(T^A,T^B)
\end{equation}
In order to serve as the gauge symmetry algebra of M2 branes world
volume theory, namely that for the  equations of motion to be
consistent with gauge symmetry and supersymmetry, the fundamental
identity need to be imposed to the 3-algebra:
\begin{eqnarray}
[T^A,T^B,[T^C,T^D,T^E]]&&=[[T^A,T^B,T^C],T^D,T^E]+[T^C,[T^A,T^B,T^D],T^E]\nonumber\\
&&+[T^C,T^D,[T^A,T^B,T^E]]
\end{eqnarray}
which extends the Jacobi identity to the 3-algebra and is equivalent
to
\begin{equation}
\label{fi2}
{f^{EFG}}_{D}{f^{ABC}}_{G}={f^{EFA}}_{G}{f^{BCG}}_{D}+{f^{EFB}}_{G}{f^{CAG}}_{D}+{f^{EFC}}_{G}{f^{ABG}}_{D}
\end{equation}
In order to derive equations of motion from the Lagrangian
description, a bi-invariant metric $h^{AB}$ on the 3-algebra is
needed which requires
\begin{equation}
{\rm Tr}([T^A,T^B,T^C],T^D)+{\rm Tr}(T^A,[T^B,T^C,T^D])=0
\end{equation}
This implies the tensor $f^{ABCD}\equiv{f^{ABC}}_{E}h^{ED}$ is
totally antisymmetric.

The BLG theory enjoys the classical conformal invariance and
$\mathcal{N}=8$ supersymmetry, which has 16 supersymmetries. The
action also has a manifest SO(8) R-symmetry that acts on the scalars
$X^{(I)}$. It has no free parameters and the structure constant of
the 3-algebra is quantized \cite{Bagger:2007vi}, which strongly
suggests the conformal invariance is exact at the quantum level. The
elegant and unique structure of the BLG theory makes it a very
compelling candidate of the multiple M2 branes theory. The BLG
theory encodes the interactions of three dimensional $\mathcal{N}=8$
multiplet. The fermionic field $\Psi$ is a Majorana spinor in $10+1$
dimensions satisfying the chirality condition
$\Gamma_{0\hat{1}\hat{2}}\Psi=-\Psi$ while the SUSY parameter
$\epsilon$ satisfies $\Gamma_{0\hat{1}\hat{2}}\epsilon=\epsilon$. As
a result, $\Psi$ has 16 real fermionic components equivalent to 8
bosonic degrees of freedom. The bosonic fields include 8 real scalar
fields $X^{(I)}_A$, (where $I=1,...8$ specifying the transverse
directions of M2 branes) and a gauge field $\mathcal{A}_\mu$ (where
$\mu=0,\hat{1},\hat{2}$ describing the longitudinal directions). In
2+1 dimensions, an ordinary gauge field has one propagating degree
of freedom. However, in the BLG theory the gauge field $\mathcal{A}$
has only a Chern-Simons term rather than canonical kinetic terms and
hence it has no propagating degree of freedom. Matter fields in the
BLG theory take values in 3-algebra, so that we have
$X^{(I)}=X^{(I)}_AT^A, \Psi=\Psi_AT^A$. The BLG Lagrangian is given
by \cite{Bagger:2007jr}
\begin{eqnarray}
\mathcal{L}&&=-\frac{1}{2}D_{\mu}X^{A(I)}D^{\mu}X^{(I)}_A+\frac{i}{2}\bar{\Psi}^A\Gamma^{\mu}D_{\mu}\Psi_A
+\frac{i}{4}f_{ABCD}\bar{\Psi}^B\Gamma^{IJ}X^{C(I)}X^{D(J)}\Psi^A\nonumber\\
&&-\frac{1}{12}(f_{ABCD}X^{A(I)}X^{B(J)}X^{C(K)})({f_{EFG}}^DX^{E(I)}X^{F(J)}X^{G(K)})\nonumber\\
&&+\frac{1}{2}\epsilon^{\mu\nu\lambda}\Big(f_{ABCD}{{\mathcal{A}}_{\mu}}^{AB}\partial_{\nu}{{\mathcal{A}}_{\lambda}}^{CD}
+\frac{2}{3}{f_{AEF}}^Gf_{CDGB}{{\mathcal{A}}_{\mu}}^{AB}{{\mathcal{A}}_{\nu}}^{CD}{{\mathcal{A}}_{\lambda}}^{EF}\Big)
\end{eqnarray}
The theory is invariant under the $\mathcal{N}=8$ SUSY
transformations:
\begin{eqnarray}\label{susy}
&&\delta X^{A(I)}=i\bar{\epsilon}\Gamma^I\Psi^A \\
&&\delta\Psi^A=D_{\mu}X^{A(I)}\Gamma^{\mu}\Gamma_I\epsilon+\frac{1}{6}X^{B(I)}X^{C(J)}X^{D(K)}f^A\,_{BCD}\Gamma_{IJK}\epsilon\ \\
&& \delta (\tilde{A}_{\mu})^A\,_B=i\bar{\epsilon}\Gamma_{\mu}
\Gamma_I X^{C(I)} \Psi^D f^A\,_{BCD}\
\end{eqnarray}
and the gauge transformations:
\begin{equation}
 \delta X^{A(I)}={{\tilde{\Lambda}}^{A}}\,_BX^{B(I)}\ , \qquad
\delta \Psi^A={\tilde{\Lambda}^A}\,_B\Psi^B\ , \qquad \delta(\tilde
{\mathcal{A}}_{\mu})^A\,_B= D_{\mu}\tilde{\Lambda}^A\,_B\ .
\end{equation}
where  $\tilde{\Lambda}^A\,_{B} = \Lambda_{MN} f^{MNA}\,_B$ and
$(\tilde{\mathcal{A}}_\mu)^A\,_{B} = ({\mathcal{A}}_\mu)_{MN}
f^{MNA}\,_B$. The gauge group is generated by the
$\tilde{\Lambda}^A\,_{B}$, while the antisymmetric $\Lambda_{MN}$
are auxiliary parameters. The gauge group is thus a subgroup of
$GL(N)$ where $N$ is the dimension of 3-algebra. If we add a metric
of signature $(N - k , k)$ on the $3$-algebra, then we can say that
the gauge group is a subgroup of $SO( N - k, k)$. The closure of the
SUSY transformations implies the equations of motion.

In most of physical theories, a positive definite metric is required
to preserve unitarity, that is the theory has positive definite
kinetic terms preventing the propagation of ghost degrees of
freedom. In the BLG theory, the positive definite metric requirement
is very strong: it was conjectured in \cite{Ho:2008bn} and then
proved in \cite{Papadopoulos:2008sk}\cite{Gauntlett:2008uf} that
there is only one non-trivial 3-algebra $\mathcal{A}_4$ satisfying
positive definite metric requirement. 3-algebra $\mathcal{A}_4$ is
4-dimensional and defined by structure constants
$f^{ABC}\,_D=\epsilon^{ABC}\,_D$, where $\epsilon^{ABCD}$ is the
4-dimensional Levi Civita symbol. New constructions are possible if
we do not require the existence of Lagrangian but only of the
equations of motion \cite {Morozov:2008cb}\cite{Gran:2008vi}, which
can be written without the help of metric in the algebra. Note that
in the Bagger-Lambert work at the level of equation of motion, the
metric is not used. The metric is needed in order to have a
Lagrangian and gauge invariant local operators. Recently there is a
breakthrough in constructing new 3-algebra
\cite{Gomis:2008uv}\cite{Benvenuti:2008bt}\cite{Ho:2008ei}. The
novel construction of 3-algebra $\mathcal{A}_{\mathcal{G}}$ is based
on an arbitrary compact and semi-simple Lie algebra $\mathcal{G}$.
These new constructions relax the requirement that the metric on the
3-algebra is positive and definite. The direction of relaxing
positive definite metric requirement has been pursued in some
earlier papers \cite{Awata:1999dz}\cite{Ho:2008bn}. Following the
convention of \cite{Gomis:2008uv}, the metric on the 3-algebra
$\mathcal{A}_{\mathcal{G}}$ is
\begin{equation}h^{AB}=\eta^{AB}\  ,  \qquad A,B=0,1,...,n+1,
\end{equation}
where $ N=n+2$ is the dimension of $\mathcal{A}_{\mathcal{G}}$ and
$\eta^{AB}$=diag(-1,1,....,1) is the Minkowski metric on 3-algebra
$\mathcal{A}_{\mathcal{G}}$. Then we split the 3-algebra indices
 ($A, B,....$ ) into ($0, a, b, ....,\phi$), where $a, b=1, ...., n$
and $\phi\equiv {n+1}$. The following form of the totally
antisymmetric structure constants satisfies the fundamental identity
(\ref{fi2}) :
\begin{equation}f^{0abc}=f^{\phi abc}=f^{abc}\  ,  \qquad f^{o\phi
ab}=f^{abcd}=0
\end{equation}
 3-index
structure constants $f^{abc}$ are the structure constants of a
compact semi-simple Lie algebra $\mathcal{G}$ and satisfy the usual
Jacobi identity. It is convenient to change the generators to the
light-cone form:
\begin{equation}
T^{\pm}=\pm T^0+T^{\phi}
\end{equation}
In this base, the metric of $\mathcal{A}_{\mathcal{G}}$ is given by
\begin{equation}
h^{+-}=h^{-+}=2\  ,  \qquad h^{++}=h^{--}=0\  ,  \qquad
h^{ab}=\delta ^{ab}\  ,  \qquad h^{\pm a}=h^{a \pm}=0
\end{equation}
and the structure constants are
\begin{eqnarray}
&f^{+abc}=-f^{a+bc}=f^{ab+c}=-f^{abc+}=2f^{abc}&\nonumber\\
&f_{-abc}=-f_{a-bc}=f_{ab-c}=-f_{abc-}=f_{abc}&\nonumber\\
&f^{-abc}=f_{+abc}=0&\nonumber
\end{eqnarray}
It is easy to see that the generator $T^{-}$ is central, viz. that
the trilinear antisymmetric product vanishes whenever $T^{-}$
appears. The Lagrangian based on $\mathcal{A}_{\mathcal{G}}$ is
given by
\begin{eqnarray}\label{BLGlag}
\mathcal{L}&=&-\frac{1}{2} h^{AB} D_{\mu}X^{(I)}_A D^{\mu}X^{(I)}_B
+\frac{i}{2} h^{AB} {\bar\Psi}_A\Gamma^{\mu}D_{\mu}\Psi_B \nonumber \\
&& - \frac{1}{12} h^{MN} f^{ABC}\,_M f^{EFG}\,_N
X^{(I)}_AX^{(J)}_BX^{(K)}_CX^{(I)}_EX^{(J)}_FX^{(K)}_G\nonumber\\
&&-\frac{i}{4}h^{DE} f^{ABC}\,_E X^{(I)}_AX^{(J)}_B
{\bar\Psi}_C\Gamma_{IJ} \Psi_D +4\epsilon^{\mu\nu\lambda} {\rm
Tr}\Big( \cB_{\lambda} (\partial_{\mu}\cA_{\nu} - [\cA_{\mu},
\cA_{\nu}]) \Big)
\end{eqnarray}
It is important to note that the Lagrangian should be derived
directly based on the new algebra
$\mathcal{A}_{\mathcal{G}}$\cite{Benvenuti:2008bt} rather than from
the result of Bagger and Lambert \cite{Bagger:2007jr}.
Bagger-Lambert Lagrangian depends on the assumption that the metric
is positive and definite. Henceforth we will call this theory
generalized Bagger-Lambert theory which is based on 3-algebra
$\mathcal{A}_{\mathcal{G}}$. The generalized Bagger-Lambert theory
does not admit any tunable coupling constant which hints that the
generalized Bagger-Lambert theory is related to M2 branes.

\section{Deformation of the Generalized BLG Theory }
In ref.\cite{Bena:2000zb}, it was argued that in the presence of a
particular background four form flux, M2 branes preserve four
supersymmetries and exhibit an SO(4) R-symmetry. Furthermore, the
flux induces a supersymmetric mass term for the world volume scalars
and fermions. It was also argued that in this background, the vacuum
of $n$ M2 branes is a state in which the scalars describe a fuzzy
three-sphere in spacetime. The M2 branes puff up so that their world
volume is of form $\mathbb{R}^{1,2}\times \widetilde{S}^3$, where
$\widetilde{S}^3$ is a fuzzy three sphere which becomes a normal
$S^3$ as $n\rightarrow\infty$. This setup provides an M-theory
analog of Myers effect which occurs for D-branes in the presence of
background fluxes \cite{Myers:1999ps}. Following this argument, the
fuzzy sphere solution of the BLG theory was found in
\cite{Bagger:2007vi} based on $\mathcal{A}_4$ algebra. The more
general deformation was considered in \cite{Gomis:2008cv} in which
two terms are added to the BLG theory. One is the mass term for all
the scalars and fermions,
\begin{equation}
\mathcal{L}_{mass}=-\frac{1}{2}{\mu}^2h^{AB}X^{(I)}_AX^{(I)}_B+\frac{i}{2}{\mu}h^{AB}{\bar
\Psi}_A \Gamma_{1234}\Psi_B\  .
\end{equation}
The other is a Myers-like SO(4)$\times$SO(4) invariant scalar
potential induced from background fluxes,
\begin{equation}
\mathcal{L}_{flux}=-\frac{1}{6}{\mu}\epsilon^{IJKL}h^{AB}[X^{(I)},X^{(J)},X^{(K)}]_AX^{(L)}_B
-\frac{1}{6}{\mu}\epsilon^{I'J'K'L'}h^{AB}[X^{(I')},X^{(J')},X^{(K')}]_AX^{(L')}_B
\end{equation}
where $I',J',K',L'=1,2,3,4$ and $ I, J, K, L=5,6,7,8$ representing
the transverse directions. It was proven in \cite{Gomis:2008cv} that
the BLG theory with the above two deformation terms remains fully
supersymmetric. Notice that this kind of supersymmetric deformation
applies to any 3-algebra with totally antisymmetric structure
constants satisfying the fundamental identity. Along this line, we
consider the deformation of the generalized BLG theory. The
Lagrangian of the deformed theory is:
\begin{equation}\label{deformedL}
\widetilde{\mathcal{L}}=\mathcal{L}+\mathcal{L}_{mass}+\mathcal{L}_{flux} \\
\end{equation}
\begin{eqnarray}\
\mathcal{L}&=&-\frac{1}{2} h^{AB} D_{\mu}X^{(I)}_A D^{\mu}X^{(I)}_B
+\frac{i}{2} h^{AB} \bar{\Psi}_A\Gamma^{\mu}D_{\mu}\Psi_B \nonumber \\
&& - \frac{1}{12} h^{MN} f^{ABC}\,_M f^{EFG}\,_N
X^{(I)}_AX^{(J)}_BX^{(K)}_CX^{(I)}_EX^{(J)}_FX^{(K)}_G\nonumber\\
&&-\frac{i}{4}h^{DE} f^{ABC}\,_E X^{(I)}_AX^{(J)}_B
{\bar\Psi}_C\Gamma_{IJ} \Psi_D +4\epsilon^{\mu\nu\lambda} {\rm
Tr}\Big( \cB_{\lambda} (\partial_{\mu}\cA_{\nu} - [\cA_{\mu},
\cA_{\nu}]) \Big)
\end{eqnarray}
\begin{equation}
\mathcal{L}_{mass}=-\frac{1}{2}{\mu}^2h^{AB}X^{(I)}_AX^{(I)}_B+\frac{i}{2}{\mu}h^{AB}{\bar
\Psi}_A \Gamma_{1234}\Psi_B
\end{equation}
\begin{equation}
\mathcal{L}_{flux}=-\frac{1}{6}{\mu}\epsilon^{IJKL}h^{AB}[X^{(I)},X^{(J)},X^{(K)}]_AX^{(L)}_B
-\frac{1}{6}{\mu}\epsilon^{I'J'K'L'}h^{AB}[X^{(I')},X^{(J')},X^{(K')}]_AX^{(L')}_B
\end{equation}
The deformed theory (\ref{deformedL}) breaks the SO(8) R-symmetry of
the BLG theory down SO(4)$\times$ SO(4). It is invariant under 16
supersymmetries. The supersymmetry transformations of the deformed
theory are given by
\begin{eqnarray}\label{susy2}
&&\delta X^{A(I)}=i\bar{\epsilon}\Gamma^I\Psi^A\nonumber \\
&&\delta\Psi^A=D_{\mu}X^{A(I)}\Gamma^{\mu}\Gamma_I\epsilon+\frac{1}{6}X^{B(I)}X^{C(J)}X^{D(K)}f^A\,_{BCD}\Gamma_{IJK}\epsilon
-\mu\Gamma_{1234}\Gamma^{I}X^{A(I)}\epsilon \ \\
&& \delta (\tilde{A}_{\mu})^A\,_B=i\bar{\epsilon}\Gamma_{\mu}
\Gamma_I X^{C(I)} \Psi^D f^A\,_{BCD}\ ,\nonumber
\end{eqnarray}
By setting $\mu\rightarrow 0$, we recover the supersymmetry
transformations (\ref{susy}) of the BLG theory. The generalized BLG
theory is invariant under another 16 non-linearly realized
supersymmetries due to the existence of the central generator $T^-$.
For the later convenience, let us clarify the notation. ($+, -, A,
B,...$) denotes the index of 3-algebra $\mathcal{A}_{\mathcal{G}}$.
($a, b,...$) denotes the index of Lie algebra ${\mathcal{G}}$.
($X^{(I)}, X^{(J)},...$) denotes the eight transverse direction and
one of them $X^{(8)}$ denotes the compactified direction. ($X^{(i)},
X^{(j)},...$) denotes the left transverse direction. Other
convention is:
\begin{eqnarray}\label{convention1}
&&\mathcal{A}_{\mu}^{-a}\equiv\mathcal{A}_{\mu}^{a}\  ,\qquad
\frac{1}{2}f^{abc}\mathcal{A}_{\mu
bc}\equiv\mathcal{B}_{\mu}^{a}\nonumber \\
&&\mathcal{A}_{\mu}=\mathcal{A}_{\mu}^aT^a\ ,  \qquad
\mathcal{B}_{\mu}=\mathcal{B}_{\mu}^{a}T^a \nonumber \\
&&\mathcal{F}_{\mu\nu}^a={\partial}_{\mu}\mathcal{A}_{\nu}^a-{\partial}_{\nu}\mathcal{A}_{\mu}^a
-2f^{abc}\mathcal{A}_{\mu}^b\mathcal{A}_{\nu}^c\nonumber \\
&&X^{\pm(I)}=\pm{X^{0(I)}}+X^{\phi(I)}\nonumber\\
&&\Psi^{\pm}=\pm\Psi^{0}+\Psi^{\phi}\nonumber\\
&&{\bar\Psi}^{\pm}=\pm{\bar\Psi}^{0}+{\bar\Psi}^{\phi}
\end{eqnarray}
\begin{eqnarray}
&&D_{\mu}X^{A(I)}={\partial}_{\mu}X^{A(I)}+f^A\
_{BCD}\mathcal{A}_{\mu}^{CD}X^{B(I)}\nonumber \\
&&D_{\mu}\Psi^{A}={\partial}_{\mu}\Psi^{A}+f^A\
_{BCD}\mathcal{A}_{\mu}^{CD}\Psi^{B}\nonumber \\
&&D_{\mu}X^{a(I)}={\partial}_{\mu}X^{a(I)}+2f^{abc}
\mathcal{A}_{\mu}^{c}X^{b(I)}-2\mathcal{B}_{\mu}^{a}X^{-(I)}\nonumber \\
&&D_{\mu}X^{+(I)}={\partial}_{\mu}X^{+(I)}+4\mathcal{B}_{\mu a}X^{a(I)}\nonumber \\
&&D_{\mu}X^{-(I)}={\partial}_{\mu}X^{-(I)}
\end{eqnarray}
Note that $D_{\mu}\Psi^a$, $D_{\mu}\Psi^+$ and $D_{\mu}\Psi^-$ have
the same forms as $D_{\mu}X^a$, $D_{\mu}X^+$ and $D_{\mu}X^-$ and
$D_{\mu}\bar{\Psi}$ has the same form as $D_{\mu}\Psi$. Using the
above convention, the BLG theory Lagrangian (\ref{BLGlag}) can be
rewritten in the $\mathcal{G}$ invariant form:
\begin{eqnarray}\ \label{BLGlag2}
\mathcal{L}=&&-\frac{1}{2}D_{\mu}X^{a(I)}D^{\mu}X^{a(I)}-\frac{1}{2}\Big({\partial}_{\mu}X^{+(I)}+4\mathcal{B}_{\mu
a}X^{a(I)}\Big){\partial}^{\mu}X^{-(I)}\nonumber \\
&&\frac{i}{2}\bar{\Psi}^a\Gamma^{\mu}D_{\mu}{\Psi}_a+\frac{i}{4}\bar{\Psi}^+\Gamma^{\mu}\partial_{\mu}{\Psi}^-
+\frac{i}{4}\bar{\Psi}^-\Gamma^{\mu}\Big(\partial_{\mu}\Psi^++4\mathcal{B}_{\mu
a}\Psi^a\Big)\nonumber \\
&&+\frac{i}{4}f_{abc}\bar{\Psi}^a\Gamma^{IJ}X^{b(I)}X^{c(J)}\Psi^-
-\frac{i}{4}f_{abc}\bar{\Psi}^-\Gamma^{IJ}X^{b(I)}X^{c(J)}\Psi^a\nonumber\\
&&+\frac{i}{2}f_{abc}\bar{\Psi}^b\Gamma^{IJ}X^{-(I)}X^{c(J)}\Psi^a
-\frac{1}{4}f_{abc}X^{a(I)}X^{b(J)}X^{-(K)}f_{ef}\
^cX^{e(I)}X^{f(J)}X^{-(K)}\nonumber
\\&&-\frac{1}{2}f_{abc}X^{a(I)}X^{b(J)}X^{-(K)}f_{ef}\
^cX^{e(I)}X^{f(K)}X^{-(J)}
+2\epsilon^{\mu\nu\lambda}\mathcal{B}_{\mu}^a\mathcal{F}_{\nu\lambda}^a
\end{eqnarray}
The mass deformation term of Lagrangian takes the form:
\begin{eqnarray}\ \label{masslag}
\mathcal{L}_{mass}=&&-\frac{1}{2}{\mu}^2X^{a(I)}X^{a(I)}-\frac{1}{2}{\mu}^2X^{+(I)}X^{-(I)}\nonumber
\\
&&+\frac{i}{2}{\mu}{\bar \Psi}^a
\Gamma_{1234}\Psi^a+\frac{i}{4}{\mu}{\bar \Psi}^+
\Gamma_{1234}\Psi^-+\frac{i}{4}{\mu}{\bar \Psi}^-
\Gamma_{1234}\Psi^+
\end{eqnarray}
The flux-inducing potential part of Lagrangian can be written:
\begin{equation} \label{fluxlag}
\mathcal{L}_{flux}=-\frac{2}{3}{\mu}\epsilon^{IJKL}f^{bcd}X_b^{(J)}X_c^{(K)}X_d^{(L)}X^{-(I)}+(IJKL\rightarrow
I'J'K'L')
\end{equation}
Notice that in (\ref{deformedL}), the mode $X^{+(I)}$ appears only
through linear form, so that it can be integrated out exactly. The
integration freezes the mode $X^{-(I)}$ to the value of the free
theory with a source-like term due to the presence of the mass term.
\begin{eqnarray}
(\partial^2-\mu^2)X^{-(I)}=0\label{constraint1}\\
(\Gamma^{\nu}\partial_{\nu}+\mu\Gamma_{1234})\Psi^{-}=0\label{constraint2}
\end{eqnarray}

This is a new feature of (\ref{deformedL}) and hints that the
negative norm states may be consistently decoupled from the physical
Hilbert space.

\section{From M2 to D2 }
In this section, we show how the deformed generalized BLG theory,
which is interpreted as a theory of coinciding membranes in
particular background, is related to the low energy description of
multiple D2 branes. The general solution to this problem is tricky
and we need to resort to Janus field theory. The reader who is
interested in this approach is referred to recent paper
\cite{Honma:2008un} and references therein. In this letter, we are
not ambitious to deal with the general solution. Instead we will fix
the problem in the specific limit, the weak background flux limit
which means that $\mu^2$ is very small. In other words $X^{-(8)}$
varies slowly along with worldvolume coordinates. In this limit,
Mukhi-Papageorgakis Higgs mechanism is easy to deal with. Following
the strategy of \cite{Mukhi:2008ux}, we make one of the scalar
fields acquire the expectation value with the restriction of
(\ref{constraint1}) and (\ref{constraint2}). Later we will make
$X^{-(8)}$ acquire the expectation value so we solve $X^{-(8)}$ only
and keep VEV of other fields vanish. The general solution is,
\begin{equation}
X^{-(8)}=Ae^{p_\mu x^\mu}+Be^{-p_\mu x^\mu}
\end{equation}
where $A$ and $B$ are integral constants and $p_\mu$ satisfies
$p^2=\mu^2$. For simplicity, we will work in the weak background
flux limit. We expect from this limit we can get some hints to the
general solution. In this limit we propose that
\begin{equation}
<X^{-(8)}>=\frac{\sqrt{2}R}{{\ell^{\frac{3}{2}}_p}}
\end{equation}
where R is the radius of a circle on which we compactify M-theory to
get type IIA string theory and $\ell_p$ is the 11 dimensional Planck
length scale. We have $R=g_s\ell_s=g^{\frac{2}{3}}_s\ell_p$ from the
string theory dualities and $g^2_{YM}=2(2\pi)^{p-2}g_s\ell^{p-3}_s$
from $D_p$ brane world volume theory, where $g_s$, $\ell_s$ are
string coupling and string length and $g_{YM}$ is $D_p$ brane world
volume theory coupling which is just  p+1 dimensional SYM coupling.
Combining the above results, we find that $<X^{-(8)}>=g_{YM}$.
Notice that\footnote {I am grateful to Zhao-Long Wang for pointing
this.} $g_{YM}$ must be function of world volume coordinates in
order to satisfy the constraint equation (\ref{constraint1}).
$\partial_\mu g_{YM}$ can be neglected compared with $g_{YM}$ in the
above limit. In order to get the super Yang-Mills theory smoothly
from its strong coupling limit, membranes worldvolume theory, we
need to take the limit $g_{YM}\gg1$. For the novel 3-algebra , the
fundamental identity implies that $\mathcal{A}_{\mathcal{G}}$
reduces to the algebra ${\mathcal{G}}\times U(1)$. It is easy to see
from structure constants $f^{Abc}\ _d$. If $A=+$, we get the general
Lie algebra ${\mathcal{G}}$. Otherwise structure constants vanish
and $U(1)$ is reduced. Notice that a VEV $<X^{-(8)}>$ does not
preserve all the supersymmetries due to the $\mu$ term in
(\ref{susy2}). This suggests that the reduced D2 branes system does
not preserve all the supersymmetries in the reduced background. Now
let us show how various terms in (\ref{BLGlag2}) (\ref{masslag})
(\ref{fluxlag}) reproduce the SYM theory in the reduced background.
Under the novel Higgs mechanism, the bosonic kinetic term of
(\ref{BLGlag2}) becomes:
\begin{eqnarray}
\mathcal{L}_{kineticB}&&=-\frac{1}{2}D_{\mu}X^{a(i)}D^{\mu}X^{a(i)}-\frac{1}{2}{D'}_{\mu}X^{a(8)}{D'}^{\mu}X^{a(8)}\nonumber\\
&&-2\mathcal{B}_{\mu}^a\mathcal{B}^{\mu
a}X^{-(8)}X^{-(8)}+2D'_{\mu}X^{a(8)}\mathcal{B}^{\mu
a}X^{-(8)}\nonumber \\
&&-\frac{1}{2}({\partial}_{\mu}X^{+(i)}+4\mathcal{B}_{\mu
a}X^{a(i)}){\partial}^{\mu}X^{-(i)}-\frac{1}{2}({\partial}_{\mu}X^{+(8)}+4\mathcal{B}_{\mu
a}X^{a(8)}){\partial}^{\mu}X^{-(8)}\nonumber\\
&&=-2g^2_{YM}\mathcal{B}_{\mu}^a\mathcal{B}^{\mu
a}+2g_{YM}D'_{\mu}X^{a(8)}\mathcal{B}^{\mu
a}\nonumber\\
&&-\frac{1}{2}D'_{\mu}X^{a(i)}D'^{\mu}X^{a(i)}-\frac{1}{2}{D'}_{\mu}X^{a(8)}{D'}^{\mu}X^{a(8)}\nonumber
\\
&&-\frac{1}{2}{\partial}_{\mu}X^{+(i)}{\partial}^{\mu}X^{-(i)}-\frac{1}{2}{\partial}_{\mu}X^{+(8)}{\partial}^{\mu}X^{-(8)}+{\rm
higher\ order}
\end{eqnarray}
The fermionic kinetic term becomes:
\begin{eqnarray}
\mathcal{L}_{kineticF}&&=\frac{i}{2}\bar{\Psi}^a\Gamma^{\mu}D'_{\mu}{\Psi}_a+\frac{i}{4}\bar{\Psi}^+\Gamma^{\mu}\partial_{\mu}{\Psi}^-
+\frac{i}{4}\bar{\Psi}^-\Gamma^{\mu}\partial_{\mu}\Psi^++{\rm
higher\ order}\nonumber \\
\end{eqnarray}
The Yukawa term of (\ref{BLGlag2}) becomes:
\begin{equation}
\mathcal{L}_{Yukawa}=\frac{i}{2}g_{YM}f_{abc}{\bar\Psi}^b\Gamma^{8i}X^{c(i)}\Psi^a+{\rm
higher\ order}
\end{equation}
The sextic potential term of (\ref{BLGlag2}) becomes:
\begin{equation}
\mathcal{L}_{potential}=-\frac{g^2_{YM}}{4}f_{abc}f_{ef}\
^cX^{a(i)}X^{b(j)}X^{e(i)}X^{f(j)}+{\rm higher\ order}
\end{equation}
The Chern-Simons terms of (\ref{BLGlag2}) is:
\begin{eqnarray}
\mathcal{L}_{CS} &&=4\epsilon^{\mu\nu\lambda}{\rm
Tr}\Big(\mathcal{B}_{\mu}(\partial_{\lambda}\mathcal{A}_{\nu}-[\mathcal{A}_{\lambda},\mathcal{A}_{\nu}])\Big)\nonumber\\
 &&=2\epsilon^{\mu\nu\lambda}{B}^a_{\mu}\mathcal{F}^a_{\nu\lambda}
\end{eqnarray}
The mass term of deformed Lagrangian (\ref{masslag}) becomes:
\begin{eqnarray}
\mathcal{L}_{mass}&&=-\frac{1}{2}g_{YM}{\mu}^2X^{+(8)}-\frac{1}{2}{\mu}^2X^{a(I)}X^{a(I)}\nonumber
\\
&&+\frac{i}{2}{\mu}{\bar \Psi}^a
\Gamma_{1234}\Psi^a+\frac{i}{4}{\mu}{\bar \Psi}^+
\Gamma_{1234}\Psi^-+\frac{i}{4}{\mu}{\bar \Psi}^-
\Gamma_{1234}\Psi^+
\end{eqnarray}
The flux-inducing term (\ref{fluxlag}) becomes:
\begin{equation}
\mathcal{L}_{flux}=-\frac{2}{3}g_{YM}\mu\epsilon^{8jkl}f^{bcd}X_b^{(j)}X_c^{(k)}X_d^{(l)}+{\rm
higher\ order}
\end{equation}
In the above expressions, we have define a new covariant derivative:
\begin{eqnarray}
D'_{\mu}X^{a(I)} &&=\partial_{\mu}X^{a(I)}-2f^a\
_{bc}X^{c(I)}\mathcal{A}_{\mu}^b
\nonumber\\
&&=\partial_{\mu}X^{a(I)}+2f^a\
_{bc}X^{b(I)}\mathcal{A}_{\mu}^c
\end{eqnarray}
Notice that the contributions of higher order terms are suppressed
in the strong coupling limit. By virtue of the nature of Lagrangian,
we find that ${B}^a_{\mu}$ is an auxiliary field appearing without
derivatives. It therefore can be eliminated via its equation of
motion. We can extracted the leading part of such solution by
neglecting the higher order terms. We therefore consider the
Lagrangian involving ${B}^a_{\mu}$:
\begin{equation}
\mathcal{L}=-2g^2_{YM}\mathcal{B}_{\mu}^a\mathcal{B}^{\mu
a}+2g_{YM}D'_{\mu}X^{a(8)}\mathcal{B}^{\mu
a}+2\epsilon^{\mu\nu\lambda}{B}^a_{\mu}\mathcal{F}^a_{\nu\lambda}+{\rm
higher\ order}
\end{equation}
and equation of motion for ${B}^a_{\mu}$:
\begin{equation}
{B}^a_{\mu}=\frac{1}{2g^2_{YM}}\epsilon_{\mu}\
^{\nu\lambda}\mathcal{F}^a_{\nu\lambda}+\frac{1}{2g_{YM}}D'_{\mu}X^{a(8)}
\end{equation}
Inserting this back into the Lagrangian (\ref{deformedL}):
\begin{eqnarray}\label{mainresult}
\widetilde{\mathcal{L}}&&=-\frac{1}{g^2_{YM}}\mathcal{F}^a_{\mu\nu}\mathcal{F}^{a
\mu\nu}-\frac{1}{2}D'_{\mu}X^{a(i)}D'^{\mu}X^{a(i)}-\frac{1}{2}\partial_{\mu}X^{-(I)}\partial^{\mu}X^{+(I)}\nonumber\\
&&+\frac{i}{2}\bar{\Psi}^a\Gamma^{\mu}D'_{\mu}{\Psi}_a+\frac{i}{4}\bar{\Psi}^+\Gamma^{\mu}\partial_{\mu}{\Psi}^-
+\frac{i}{4}\bar{\Psi}^-\Gamma^{\mu}\partial_{\mu}\Psi^+\nonumber\\
&&+\frac{i}{2}g_{YM}f_{abc}{\bar\Psi}^b\Gamma^{8i}X^{c(i)}\Psi^a-\frac{g^2_{YM}}{4}f_{abc}f_{ef}\
^cX^{a(i)}X^{b(j)}X^{e(i)}X^{f(j)}\nonumber\\
&&-\frac{1}{2}g_{YM}{\mu}^2X^{+(8)}-\frac{1}{2}{\mu}^2X^{a(I)}X^{a(I)}+\frac{i}{2}{\mu}{\bar
\Psi}^a \Gamma_{1234}\Psi^a+\frac{i}{4}{\mu}{\bar \Psi}^+
\Gamma_{1234}\Psi^-\nonumber\\
&&+\frac{i}{4}{\mu}{\bar \Psi}^- \Gamma_{1234}\Psi^+
-\frac{2}{3}g_{YM}\mu\epsilon^{8jkl}f^{bcd}X_b^{(j)}X_c^{(k)}X_d^{(l)}+{\rm
higher\ order}
\end{eqnarray}
A re-definition $\mathcal{A}\rightarrow\frac{1}{2}\mathcal{A}$ leads
to
\begin{eqnarray}
&&D'_{\mu}X^{a(i)}\rightarrow\partial_{\mu}X^{a(i)}+f^a\
_{bc}X^{b(i)}\mathcal{A}_{\mu}^c\equiv D_{\mu}X^{a(i)}\\
&&\mathcal{F}^a_{\mu\nu}\rightarrow\frac{1}{2}({\partial}_{\mu}\mathcal{A}_{\nu}^a-{\partial}_{\nu}\mathcal{A}_{\mu}^a
-f^{abc}\mathcal{A}_{\mu}^b\mathcal{A}_{\nu}^c)\equiv\frac{1}{2}{F}^a_{\mu\nu}
\end{eqnarray}
Notice that the Lagrangian (\ref{mainresult}) can be written in
form:
\begin{equation}
\widetilde{\mathcal{L}}=\widetilde{\mathcal{L}}_{decoupled}+\widetilde{\mathcal{L}}_{coupled}
\end{equation}
For the coupled part, we re-scale the fields as $(X,
\Psi)\rightarrow (X/{g_{YM}},{\Psi}/{g_{YM}})$, and then the
Lagrangian becomes:
\begin{equation}
\widetilde{\mathcal{L}}_{coupled}=\frac{1}{g^2_{YM}}\widetilde{\mathcal{L}}_{0}+\mathcal{O}(\frac{1}{g^3_{YM}})
\end{equation}
\begin{eqnarray}
&&\widetilde{\mathcal{L}}_0\nonumber
\\&&=-\frac{1}{4}{F}^a_{\mu\nu}{F}^{a\mu\nu}-\frac{1}{2}D_{\mu}X^{a(i)}D^{\mu}X^{a(i)}
+\frac{i}{2}\bar{\Psi}^a\Gamma^{\mu}D_{\mu}{\Psi}_a\nonumber\\
&&+\frac{i}{2}f_{abc}{\bar\Psi}^b\Gamma^{8i}X^{c(i)}\Psi^a-\frac{1}{4}f_{abc}f_{ef}\
^cX^{a(i)}X^{b(j)}X^{e(i)}X^{f(j)}\nonumber\\
&&-\frac{1}{2}{\mu}^2X^{a(i)}X^{a(i)}+\frac{i}{2}{\mu}{\bar \Psi}^a
\Gamma_{1234}\Psi^a-\frac{2}{3}\mu\epsilon^{8jkl}f^{bcd}X_b^{(j)}X_c^{(k)}X_d^{(l)}
\end{eqnarray}
This is just the SU(N) SYM theory with the mass term for scalars and
fermions and an induced scalar potential term which is just the
Myers term for D-branes in the presence of background fluxes. For
the decoupled part, we have:
\begin{eqnarray}
&&\widetilde{\mathcal{L}}_{decoupled}\nonumber\\
&&=-\frac{1}{2}\partial_{\mu}X^{-(I)}\partial^{\mu}X^{+(I)}+\frac{i}{4}\bar{\Psi}^+\Gamma^{\mu}\partial_{\mu}{\Psi}^-
+\frac{i}{4}\bar{\Psi}^-\Gamma^{\mu}\partial_{\mu}\Psi^+\nonumber\\
&&-\frac{1}{2}g_{YM}{\mu}^2X^{+(8)}-\frac{1}{2}{\mu}^2X^{a(8)}X^{a(8)}+\frac{i}{4}{\mu}{\bar
\Psi}^+ \Gamma_{1234}\Psi^-+\frac{i}{4}{\mu}{\bar \Psi}^-
\Gamma_{1234}\Psi^+\nonumber \\
&&=-\frac{1}{2}\partial_{\mu}X^{\phi(I)}\partial^{\mu}X^{\phi(I)}+\frac{1}{2}\partial_{\mu}X^{0(I)}\partial^{\mu}X^{0(I)}
+\frac{i}{2}\bar{\Psi}^{\phi}\Gamma^{\mu}\partial_{\mu}{\Psi}^{\phi}-\frac{i}{2}\bar{\Psi}^{0}\Gamma^{\mu}\partial_{\mu}{\Psi}^{0}\label{decoupled1}\\
&&+\frac{i}{2}\mu\bar{\Psi}^{\phi}\Gamma_{1234}{\Psi}^{\phi}-\frac{i}{2}\mu\bar{\Psi}^0\Gamma_{1234}{\Psi}^0
-\frac{1}{2}g_{YM}{\mu}^2(X^{0(8)}+X^{\phi(8)})-\frac{1}{2}{\mu}^2X^{a(8)}X^{a(8)}\label{decoupled2}
\end{eqnarray}
where we have used (\ref{convention1}) to rewrite the above
expression. Notice that the modes $X^{\phi(8)}$ and $X^{0(8)}$ are
free and can be dualised to two U(1) gauge fields by Abelian
duality. Thus (\ref{decoupled1}) of the decoupled Lagrangian tell us
that there are two copies of U(1) gauge theory. One U(1) theory is
normal, and the other is ghost. The rest part of decoupled
Lagrangian (\ref{decoupled2}) adds some extra terms to the two
copies of U(1) theories. The first two terms in (\ref{decoupled2})
are the mass terms for two U(1) theories respectively. The second
term in (\ref{decoupled2}) contains two source-like terms for two
U(1) theories respectively. The last term in (\ref{decoupled2}) is a
non-propagating term and does not affect the dynamics of the theory.
Therefore we find that in the limit we take above, the ghost
Lagrangian is completely decoupled from the SU(N)$\times$U(1) and it
does not affect the unitary.

\section{Discussion}
In this letter, we start from the multiple M2 branes world volume
theory which is based on a novel 3-algebra without the positive
definite metric assumption. Then we deform the world volume theory
with background fluxes and the induced mass terms. The deformed
theory is supersymmetric and without tunable coupling constant.
These properties are expected for the field theory living on
multiple M2 branes. Frankly speaking, the approach of this letter is
just an approximate approach of the complete solution. The
motivation of this letter is that the study of  M2 branes model may
give us some hints about the Chern-Simons like term of multiple M2
branes in flux background. My colleagues have done some research
work on this topic \cite{Li:2008eza}. We expect that the result of
our approximate approach will give some hints to their research.  In
\cite{Sethi:1997sw}\cite{Banks:1996my}, the theory of coincident M2
branes on $R\times T^2$ was argued to provide the Matrix theory
description of Type IIB string theory, and the correspondence in
some limits have been tested \cite{Gomis:2008cv}. The study of the
deformed M2 branes theory may lead new insights in understanding the
stringy physics in Type IIB Matrix theory. We have shown that when
one component of the scalar fields develops a VEV, the ensuing Higgs
mechanism produces a strong coupled SYM on arbitrary number D2
branes in reduced background. It is interesting to find that after
Higgsing, the ghost part of original M2 branes theory is
consistently decoupled. Though there are hints suggesting that the
negative norm states can be consistently decoupled from the physical
Hilbert space, the complete analysis remains to be uncovered.

\section*{Acknowledgments}
I would like to acknowledge useful conversations with Miao Li, Tower
Wang and Zhao-Long Wang. I apologize to everybody who I did not cite
in the previous version of this paper.


\begin{thebibliography}{99}

\bibitem{Bagger:2006sk}
  J.~Bagger and N.~Lambert,
  ``Modeling multiple M2's,''
  Phys.\ Rev.\  D {\bf 75} (2007) 045020
  [arXiv:hep-th/0611108].
  %%CITATION = PHRVA,D75,045020;%%

\bibitem{Bagger:2007jr}
  J.~Bagger and N.~Lambert,
  ``Gauge Symmetry and Supersymmetry of Multiple M2-Branes,''
  Phys.\ Rev.\  D {\bf 77} (2008) 065008
  [arXiv:0711.0955 [hep-th]].
  %%CITATION = PHRVA,D77,065008;%%

\bibitem{Bagger:2007vi}
  J.~Bagger and N.~Lambert,
  ``Comments On Multiple M2-branes,''
  JHEP {\bf 0802} (2008) 105
  [arXiv:0712.3738 [hep-th]].
  %%CITATION = JHEPA,0802,105;%%

  \bibitem{Gustavsson:2007vu}
  A.~Gustavsson,
  ``Algebraic structures on parallel M2-branes,''
  arXiv:0709.1260 [hep-th].
  %%CITATION = ARXIV:0709.1260;%%

\bibitem{Gustavsson:2008dy}
  A.~Gustavsson,
  ``Selfdual strings and loop space Nahm equations,''
  arXiv:0802.3456 [hep-th].
  %%CITATION = ARXIV:0802.3456;%%

\bibitem{Gomis:2008uv}
  J.~Gomis, G.~Milanesi and J.~G.~Russo,
  ``Bagger-Lambert Theory for General Lie Algebras,''
  arXiv:0805.1012 [hep-th].
  %%CITATION = ARXIV:0805.1012;%%

\bibitem{Benvenuti:2008bt}
  S.~Benvenuti, D.~Rodriguez-Gomez, E.~Tonni and H.~Verlinde,
  ``N=8 superconformal gauge theories and M2 branes,''
  arXiv:0805.1087 [hep-th].
  %%CITATION = ARXIV:0805.1087;%%

  \bibitem{Ho:2008ei}
  P.~M.~Ho, Y.~Imamura and Y.~Matsuo,
``M2 to D2 revisited,''
  arXiv:0805.1202 [hep-th].
  %%CITATION = ARXIV:0805.1202;%%. Henceforth we

\bibitem{Maldacena:1997re}
  J.~M.~Maldacena,
``The large N limit of superconformal field theories and
supergravity,''
  Adv.\ Theor.\ Math.\ Phys.\  {\bf 2} (1998) 231
  [Int.\ J.\ Theor.\ Phys.\  {\bf 38} (1999) 1113]
  [arXiv:hep-th/9711200].
  %%CITATION = IJTPB,38,1113;%%

  \bibitem{Mukhi:2008ux}
  S.~Mukhi and C.~Papageorgakis,
  ``M2 to D2,''
  arXiv:0803.3218 [hep-th].
  %%CITATION = ARXIV:0803.3218;%%

\bibitem{Gomis:2008cv}
  J.~Gomis, A.~J.~Salim and F.~Passerini,
  ``Matrix Theory of Type IIB Plane Wave from Membranes,''
  arXiv:0804.2186 [hep-th].
  %%CITATION = ARXIV:0804.2186;%%


\bibitem{Hosomichi:2008qk}
  K.~Hosomichi, K.~M.~Lee and S.~Lee,
  ``Mass-Deformed Bagger-Lambert Theory and its BPS Objects,''
  arXiv:0804.2519 [hep-th].
  %%CITATION = ARXIV:0804.2519;%%

\bibitem{Honma:2008un}
  Y.~Honma, S.~Iso, Y.~Sumitomo and S.~Zhang,
  ``Janus field theories from multiple M2 branes,''
  arXiv:0805.1895 [hep-th].
  %%CITATION = ARXIV:0805.1895;%%

\bibitem{Bandres:2008vf}
  M.~A.~Bandres, A.~E.~Lipstein and J.~H.~Schwarz,
``N = 8 Superconformal Chern--Simons Theories,''
  JHEP {\bf 0805}, 025 (2008)
  [arXiv:0803.3242 [hep-th]];
  %%CITATION = JHEPA,0805,025;%%

\bibitem{Berman:2008be}
  D.~S.~Berman, L.~C.~Tadrowski and D.~C.~Thompson,
``Aspects of Multiple Membranes,''
  arXiv:0803.3611 [hep-th];
  %%CITATION = ARXIV:0803.3611;%%

\bibitem{VanRaamsdonk:2008ft}
  M.~Van Raamsdonk,
``Comments on the Bagger-Lambert theory and multiple M2-branes,''
  arXiv:0803.3803 [hep-th];
  %%CITATION = ARXIV:0803.3803;%%

\bibitem{Lambert:2008et}
N.~Lambert and D.~Tong, ``Membranes on an Orbifold,''
  arXiv:0804.1114 [hep-th];
  %%CITATION = ARXIV:0804.1114;%%

\bibitem{Distler:2008mk}
  J.~Distler, S.~Mukhi, C.~Papageorgakis and M.~Van Raamsdonk,
``M2-branes on M-folds,''
  JHEP {\bf 0805}, 038 (2008)
  [arXiv:0804.1256 [hep-th]];
  %%CITATION = JHEPA,0805,038;%%

\bibitem{Bergshoeff:2008cz}
   E.~A.~Bergshoeff, M.~de Roo and O.~Hohm,
  ``Multiple M2-branes and the Embedding Tensor,''
  arXiv:0804.2201 [hep-th];
  %%CITATION = ARXIV:0804.2201;%%

\bibitem{Shimada:2008xy}
 H.~Shimada,
``$\beta$-deformation for matrix model of M-theory,''
  arXiv:0804.3236 [hep-th];
  %%CITATION = ARXIV:0804.3236;%%

\bibitem{Ho:2008nn}
  P.~M.~Ho and Y.~Matsuo,
  ``M5 from M2,''
  arXiv:0804.3629 [hep-th];
  %%CITATION = ARXIV:0804.3629;%%

\bibitem{Morozov:2008rc}
   A.~Morozov,
  ``From Simplified BLG Action to the First-Quantized M-Theory,''
  arXiv:0805.1703 [hep-th];
  %%CITATION = ARXIV:0805.1703;%%

\bibitem{Fuji:2008yj}
  H.~Fuji, S.~Terashima and M.~Yamazaki,
  ``A New N=4 Membrane Action via Orbifold,''
  arXiv:0805.1997 [hep-th];
  %%CITATION = ARXIV:0805.1997;%%

\bibitem{Ho:2008ve}
  P.~M.~Ho, Y.~Imamura, Y.~Matsuo and S.~Shiba,
  ``M5-brane in three-form flux and multiple M2-branes,''
  arXiv:0805.2898 [hep-th].
  %%CITATION = ARXIV:0805.2898;%%

\bibitem{Krishnan:2008zm}
C.~Krishnan and C.~Maccaferri, ``Membranes on Calibrations,''
arXiv:0805.3125 [hep-th].
%%CITATION = ARXIV:0805.3125;%%


\bibitem{Bena:2000zb}
  I.~Bena,
``The M-theory dual of a 3 dimensional theory with reduced
supersymmetry,''
  Phys.\ Rev.\  D {\bf 62}, 126006 (2000)
  [arXiv:hep-th/0004142].
  %%CITATION = PHRVA,D62,126006;%%,


\bibitem{Myers:1999ps}
  R.~C.~Myers,
``Dielectric-branes,''
  JHEP {\bf 9912}, 022 (1999)
  [arXiv:hep-th/9910053].
  %%CITATION = JHEPA,9912,022;%%.

\bibitem{Ho:2008bn}
  P.~M.~Ho, R.~C.~Hou and Y.~Matsuo,
  ``Lie 3-Algebra and Multiple M2-branes,''
  arXiv:0804.2110 [hep-th].
  %%CITATION = ARXIV:0804.2110;%%


 \bibitem{Papadopoulos:2008sk}
  G.~Papadopoulos,
  ``M2-branes, 3-Lie Algebras and Plucker relations,''
  arXiv:0804.2662 [hep-th].
  %%CITATION = ARXIV:0804.2662;%%

\bibitem{Gauntlett:2008uf}
  J.~P.~Gauntlett and J.~B.~Gutowski,
  ``Constraining Maximally Supersymmetric Membrane Actions,''
  arXiv:0804.3078 [hep-th].
  %%CITATION = ARXIV:0804.3078;%%



\bibitem{Morozov:2008cb}
 A.~Morozov,
  ``On the Problem of Multiple M2 Branes,''
  arXiv:0804.0913 [hep-th].
  %%CITATION = ARXIV:0804.0913;%%

  \bibitem{Gran:2008vi}
  U.~Gran, B.~E.~W.~Nilsson and C.~Petersson,
  ``On relating multiple M2 and D2-branes,''
  arXiv:0804.1784 [hep-th].
  %%CITATION = ARXIV:0804.1784;%%

  \bibitem{Awata:1999dz}
  H.~Awata, M.~Li, D.~Minic and T.~Yoneya,
  ``On the quantization of Nambu brackets,''
  JHEP {\bf 0102} (2001) 013
  [arXiv:hep-th/9906248].
  %%CITATION = JHEPA,0102,013;%%


\bibitem{Sethi:1997sw}
  S.~Sethi and L.~Susskind,
  ``Rotational invariance in the M(atrix) formulation of type IIB theory,''
  Phys.\ Lett.\  B {\bf 400}, 265 (1997)
  [arXiv:hep-th/9702101].
  %%CITATION = PHLTA,B400,265;%%

\bibitem{Banks:1996my}
  T.~Banks and N.~Seiberg,
``Strings from matrices,''
  Nucl.\ Phys.\  B {\bf 497}, 41 (1997)
  [arXiv:hep-th/9702187].
  %%CITATION = NUPHA,B497,41;%%


\bibitem{Li:2008eza}
  M.~Li and T.~Wang,
  ``M2-branes Coupled to Antisymmetric Fluxes,''
  JHEP {\bf 0807}, 093 (2008)
  [arXiv:0805.3427 [hep-th]].
  %%CITATION = JHEPA,0807,093;%%
\end{thebibliography}
\end{document}